\documentclass[fleqn,usenatbib]{mnras}

\usepackage{newtxtext,newtxmath}
\usepackage[normalem]{ulem}

\usepackage[T1]{fontenc}

\DeclareRobustCommand{\VAN}[3]{#2}
\let\VANthebibliography\thebibliography
\def\thebibliography{\DeclareRobustCommand{\VAN}[3]{##3}\VANthebibliography}

\usepackage{graphicx}	
\usepackage{amsmath}	
\usepackage{caption}
\usepackage{subcaption}

\newcommand{\cmg}{\text{cm}^2\,\text{g}^{-1}}

\makeatletter
\newcommand*{\rom}[1]{\expandafter\@slowromancap\romannumeral #1@}
\makeatother

\title[subhalo core collapse]{The abundance of core--collapsed subhalos in SIDM: insights from structure formation in $\Lambda$CDM}

\author[N. Shah \& S. Adhikari]{
Neev Shah,$^{1}$\thanks{E-mail: neev.shah@students.iiserpune.ac.in}
Susmita Adhikari$^{1}$
\\
$^{1}$Department of Physics, Indian Institute of Science Education and Research Pune, Maharashtra, 411045, India\\
}

\date{Accepted XXX. Received YYY; in original form ZZZ}

\pubyear{2022}

\begin{document}
\label{firstpage}
\pagerange{\pageref{firstpage}--\pageref{lastpage}}
\maketitle

\begin{abstract}
Dark matter halos can  enter a phase of gravothermal core--collapse in the presence of self-interactions. This phase that follows a core--expansion phase is thought to be subdominant due to the long time-scales involved.  However, it has been shown that the collapse can be accelerated in tidal environments particularly for halos that are centrally concentrated. Cosmological simulations in $\Lambda$CDM give us the full distribution of satellite orbits and halo profiles in the universe. We use properties of the orbits and profiles of subhalos from simulations to estimate the fraction of the subhalos in different host halo environments, ranging from the Large Magellanic cloud(LMC)--like hosts to clusters, that are in the core--collapse phase. We use fluid simulations of self--interacting dark matter (SIDM) to evolve subhalos in their hosts including the effect of tidal truncation at the time of their pericenter crossing. We find that for parameters that allow the interaction cross-section to be high at dwarf scales, at least $10~\%$ of all subhalos are expected to have intrinsically collapsed within Hubble time up to the group mass host scales. This fraction increases significantly, becoming at least 20$\%$ when tidal interactions are considered. To identify these objects we find that we either need to measure their densities at very small radial scales, where the subhalos show a bimodal distribution of densities,  or alternatively we need to measure the slopes of their inner density profiles near the scale radius, which are much steeper than NFW slopes expected in cold dark matter halos. Current measurements of central slopes of classical dwarfs do not show a preference for collapsed objects, however this is consistent with an SIDM scenario where the classical dwarfs are expected to be in a cored phase. 
\end{abstract}




\section{Introduction}

Dark Matter self-interactions arise naturally in physics beyond the standard model. While they were originally invoked to solve small scale problems \citep{Klypin9901240, Lazar190708841, Bullock170704256}, over years advancement of observations and simulations have opened up a wider range of phenomenology \citep{Tulin:2017ara, Buckley:2017ijx, 2022arXiv220710638A}. Dark matter self-interactions provide an avenue for exchange of momentum between dark matter particles affecting the structure and distribution of dark matter particularly within dark matter halos.

In the cold dark matter paradigm, simulations have shown that the mass density profile  of dark matter halos are expected to follow NFW. Introduction of self-interaction allows the transfer of heat from the outskirts of the halo where the velocity dispersion is high to the inner regions that are cooler in a CDM halo, with time driving the region within the scale radius to have an isothermal velocity distribution. This redistribution of energy manifests itself as a core in the density profile near the centre as opposed to the cuspy or steep rise expected in the cold dark matter scenario \citep{Spergel:1999mh}. The core expands till the inner region is completely thermalized with the surrounding halo. The evolution of the inner density profile in SIDM is primarily determined by two timescales, namely, the mean collision time of the dark matter particles with itself and the gravitational dynamical time at the location within the host dark matter halo. As the core expands and thermalizes a negative energy gradient is eventually established, the random motions in the core is no longer able to support the gravity and the core expansion phase ends, leading eventually to core collapse \citep{Balberg0110561}, a process similar to that which occurs in globular clusters \citep{1980MNRAS.191..483L}. In typical massive halos of the size of Milky Way or galaxy clusters, with currently allowed cross-sections, the timescales for the core to collapse is usually longer than a Hubble time for elastic hard-sphere like scattering, therefore the primary effect of SIDM is to be observed as a cored density profile (at least in the absence of baryons). 

However, several recent works have shown that the environment of the halo can influence the process of core collapse significantly. \cite{Nishikawa190100499} showed that particularly for halos that live in strong tidal environments, like subhalos, can have their core collapse significantly accelerated compared to halos of the same mass that live in isolated environments. While \cite{Nishikawa190100499} was primarily based on a semi-analytic treatment of dark matter halos, \cite{Zeng:2021ldo} performed simulations to show that the environmental effects can have subtle impact on the timescales of collapse, in fact depending on the orbital phase of the halo either a delay or acceleration in core collapse can take place. In particular post pericenter passage the effect of tidal heating and SIDM evaporation can play a role in slowing down the core--collapse. 

Primarily the timescale for collapse depends on the inner density of the dark matter halo, parameterized often as the concentration of the halo and naturally the cross-section of interaction between dark matter particles. While the cross-section has been constrained to be less than $\sim \cmg$ in the massive cluster range by observations of the Bullet cluster \citep{Markevitch:2003at}, the cross-section can in principle be much larger at lower halo mass scales where the internal velocity dispersion is much lower than Bullet cluster like systems.  Such velocity dependent cross-sections are motivated also from particle physics, where the dark matter interactions mediated by light mediators like light scalars or dark photons give rise to Yukawa like velocity dependent interactions \citep{Buckley:2009in, Loeb:2010gj,  Tulin:2012wi}. At high velocity scales there is a Rutherford like $v^{-4}$ fall off, while the cross-section is constant at low velocities. The higher cross-sections at lower velocities compounded with the fact that the concentration of halos in CDM increases for low masses make low mass halos intriguing candidates for testing the nature of dark matter and in particular core--collapse. These halos are also more likely than larger halos to live in tidal environments, within the virial radius of other halos, orbiting in their potential.

Some recent observations have pointed at intriguing behaviour of satellite systems that can be explained by core collapse in tidal environments, for e.g. \cite{Kaplinghat190404939} find an anti-correlation between pericenter distances of Milky Way dwarfs with their inferred central densities, \cite{Minor:2020hic} detected a dark substructure using strong lensing that seems to have a slope higher than that expected from NFW.  

In this paper we use N-Body cosmological simulations of cold-dark matter halos to employ the full statistical distribution of properties of subhalos with their complete cosmological evolution to study the candidates that are likely to undergo core--collapse in the universe. We extract properties of subhalos from CDM simulations at early times and use them to evolve fluid simulations to predict their behaviour in tidal environments. We use zoom--in simulations of halos ranging from the LMC mass-scale to the cluster mass scale to get the full distribution of orbital times and profiles of all subhalos and study the statistical distribution of properties that make them prone to core collapse.  

This paper is organized as follows, in Section \ref{sec:sims} we describe the simulations used to analyze the structural properties of the dark matter subhalos  in  Section \ref{sec:collapse_times} we describe the semi analytical models to predict the collapse time in isolated environments and in tidal environments. In the same section we present our results for the distribution of collapse times in different host halos. In Section \ref{sec:discussion} we discuss the implications of our results.

\section{Simulations}
\label{sec:sims}

In this work, we use the Symphony suite of simulations \citep{Nadler:2022dvo}, which are a compilation of 262 cosmological, cold dark matter--only zoom--in simulations spanning host halo masses of $10^{11}M_\odot$(LMC-sized) to $10^{15}M_\odot$(Cluster-sized). The Symphony suites provide extensive properties of the subhalos that have been simulated at high resolution. The simulations are described in more detail in \citep{Nadler:2022dvo}. The full suite includes 39 LMC--mass, 45 Milky Way--mass \citep{Mao:2015yua}, 49 Group--mass, 33 Low--mass Clusters \citep{Bhattacharyya:2021vyd} and 96 higher mass Clusters \citep{Wu:2013rfa}  host halos. The particle mass resolutions are $5 \times 10^{4}M_\odot$ for the LMC--mass host, $4 \times 10^{5}M_\odot$ for the Milky Way--mass host, $3.3 \times 10^{6}M_\odot$ in the Group host and $1.8 \times 10^{8}M_\odot$ and $2.2 \times 10^{8}M_\odot$ for the L-cluster and Cluster simulations respectively.

The halos and subhalos in the simulation were found using the Rockstar halo finder \citep{Rockstar}. For every simulation we find all the halos that have ever been located inside the host virial radius at any snapshot. These form our complete subhalo population. Like \citep{Nadler:2022dvo} we only consider subhalos that survive at $z=0$ and roughly have more than 300 particles at accretion.

\section{Evaluation of Collapse time}
\label{sec:collapse_times}

\begin{figure*}
    \centering
    \includegraphics[width=\textwidth]{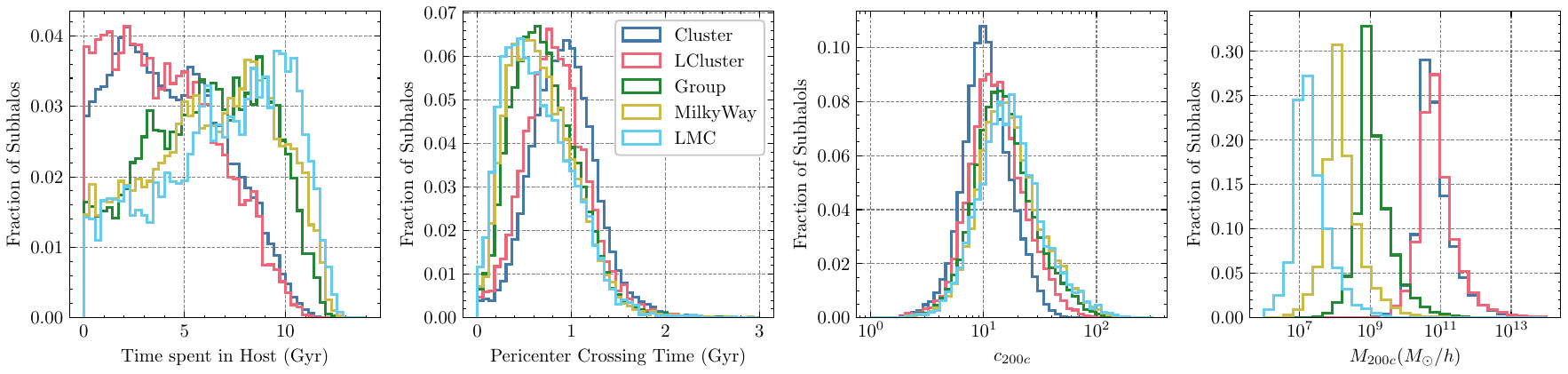}
    \caption{The four panels show the distributions of various properties of the subhalos that survive at $z=0$ across all the suites. The top left panel shows the distribution of infall times of the subhalos. Subhalos in Milky Way and LMC suites have lived inside their hosts longer on average compared to subhalos in clusters and groups. The second panel shows the distribution of their first-pericenter times. The third panel shows the concentration of the subhalos at infall and the last panel shows the masses at infall. On average the concentration of cluster subhalos is lower than subhalos in low mass hosts and their masses are higher. In this panel we plot the entire subhalo population without imposing a cut of 300 particles at accretion.}
    \label{fig:subhalo-properties}
\end{figure*}

Our goal is to evaluate the distribution of collapse times for the full subhalo population above the resolution limit of a given Symphony host dark matter halo. We track the evolution of all the dark matter halos that have ever fallen into the main host halo and have survived as a subhalo till the current time at $z=0$. 
Since we have access to the entire merger history of the subhalo we extract its properties at infall or accretion, this is defined as the time when they first cross the virial radius of the host halo. The distribution of their masses, concentrations, accretion time and time taken to reach pericenter are shown in Fig. \ref{fig:subhalo-properties}. The infall masses and internal velocity dispersions are used to estimate the interaction cross-section at the velocity scale of the subhalo. We make the choice to use infall properties to avoid the complications and uncertainties in subhalo profile parameters that emerge when they fall into a host due to physical tidal disruption and artificial disruption \citep{VandenBosch171105276}. In particular it is also known that galaxy luminosities tend to trace peak and infall properties of subhalos in CDM simulations \citep{Reddick:2012qy}. 

In this work we assume that the same subhalos survive to $z=0$ in the SIDM scenario as in CDM, this is true when the cross--section at the host velocity scale is small and subhalo evaporation is suppressed (which is true in our fiducial models). It is possible, however, that the survival probability may change due to the change in the internal profiles and the consequent difference in tidal effects and there may be some discrepancies in detail.  Our work predicts the probability of core--collapse of the population that falls into the host, the survival probability should ideally be folded into the detectability of such systems. \footnote{The number of completely disrupted subhalos in SIDM simulations is not large as reported in \cite{Dooley160308919}, however \cite{Nadler200108754} find significantly more disruption. The details of the implementation of the SIDM interactions, which are different in the two works, may have an effect on the inferred disruption probability, and we do not fold that in this work as we restrict ourselves to CDM properties.}

For our analysis we use a velocity--dependent differential self--interaction cross--section. This has a Yukawa-like $1/v^4$ fall off at high velocities and an isotropic, constant cross--section at low velocities. The angular cross--section is given by,

\begin{equation}
\frac{d\sigma}{d\Omega}=\frac{\sigma_0}{4\pi}\frac{1}{\left[1+\frac{v^2}{w^2}\sin^2\frac{\theta}{2}\right]^2}
\label{eq:sigma_t}
\end{equation}
where $v$ is the relative velocity between the particles, $w$ is a parameter that sets the transition between the $v^{-4}$ and isotropic regime at a given velocity scale. The overall normalization constant $\sigma_0$ is set by requiring that the momentum transfer cross-section, $\sigma_T/m=1$ at a chosen velocity scale, for details see \citep{Kummer170604794, Banerjee:2019bjp}. 

A given dark matter halo can be associated with a total virial mass and a corresponding characteristic circular velocity, $V_{\rm max}$, if an NFW profile is assumed. The velocity dependence of the self--interaction cross--section therefore often manifests itself as a characteristic cross--section at every halo mass scale.  However, one can in principle not simply use the $V_{\rm max}$ for a given halo but instead account for the velocity distribution of the particles within it and evaluate an effective cross--section at a given halo mass or $V_{\rm max}$. The total effective cross-section for a halo assuming an NFW distribution of its particles is given by \citep{Yang:2022hkm,2023MNRAS.523.4786O},

\begin{equation}{\label{effective-sigma}}
    \sigma_{\textrm{eff}}/m = \frac{1}{512v_{\textrm{eff}}^8}\int v^2 dvd{\cos}\theta \frac{d\sigma}{d{\cos}\theta}v^5{\sin}^2\theta ~ 
 \textrm{exp}\left(-\frac{v^2}{4v^2_{\textrm{eff}}}\right),
\end{equation}
\begin{figure}
\centering
     \includegraphics[width=0.9\columnwidth]{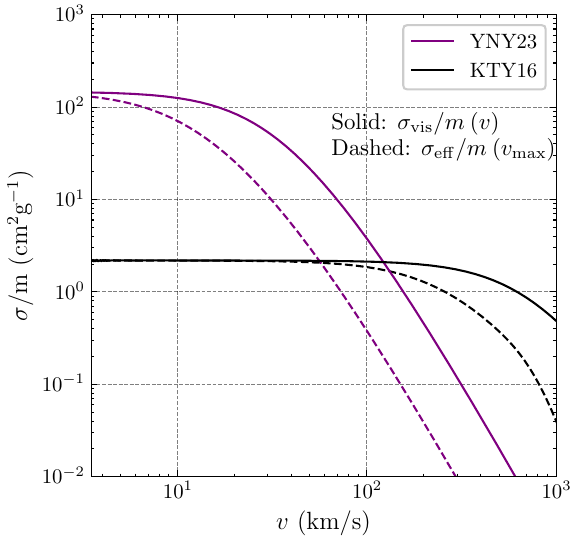}
    \caption{The dependence of the effective cross sections on the halo maximum circular velocity (essentially the dependence on halo mass). The effective cross section is evaluated by using equation \ref{effective-sigma}, that integrates over the internal velocity distribution of the halo at a given $V_{\rm max}$ for 2 models, YNY23 and KTY16. We also show the viscosity cross section as a function of velocity for the 2 models.}
    \label{fig:sigma-v}
\end{figure}
where the Gaussian is an approximation for the velocity distribution of particles in the virialized dark matter halo. The parameter $v_{\textrm{eff}} \approx 0.64~V_{\textrm{max}}$ for a halo with a NFW profile where $V_{\rm max}$ is the maximum internal circular velocity of the halo. Fig. \ref{fig:sigma-v} shows the effective cross-section as a function of halo $V_{\rm max}$ as well as the viscosity cross section as a function of velocity. Each subhalo is assigned a different effective cross section using Eqn. \ref{effective-sigma}, based on their maximum internal circular velocity, $V_{\rm max}$ at infall. In this work, following \citep{2023ApJ...949...67Y}, we set $w = 24.33$ $\rm{km}/\rm{s}$ and $\sigma_0 = 147.1$ $\rm cm^2/\rm{g}$, which is equivalent to the "vd100" model in \citep{2021MNRAS.505.5327T}. Following \citep{2023arXiv230601830N}, this is henceforth referred to as the YNY23 model. This model sets the momentum transfer cross section ($\sigma_T/m$) to be $100$ $\rm cm^2/\rm g$ at
$v = 14$ $\rm km/\rm s$ and $2$ $\rm cm^2/\rm g$
at $v = 100$ $\rm km/\rm s$ and is explored as it allows for large interaction cross-sections at low velocity scales which is relatively unconstrained. We use this as the fiducial model. For comparison we also show results (where relevant) using the current constraints on the velocity dependent model using galaxy and cluster data from \citep{Kaplinghat:2015aga}  (KTY16 model from hereon), this sets $w = 600$ $\rm{km}/\rm{s}$ and $\sigma_0 = 2.19$ $\rm cm^2/\rm{g}$. If a model name is not specified, the YNY23 model was used for the corresponding figure. Using the effective SIDM cross-section we describe in the next subsections two ways to estimate the collapse time of a subhalo. 

\subsection{Intrinsic collapse times}

We estimate the collapse times of the subhalos that have survived to redshift $z=0$, firstly, using the intrinsic collapse time; i.e. the  time that an isolated halo of a given mass will take to reach the core collapse phase starting from an NFW--like profile. This can be estimated analytically by treating the evolution of the dark matter in a halo in the presence of self-interactions as a gravothermal fluid \citep{Gnedin04, Balberg0110561, Koda11013097, Pollack:2014rja}. The full treatment will be described in section \ref{sec:fluid}. It is found that for completely elastic scattering the time for collapse (in the long-mean-free-path regime) can be estimated by a simple scaling relation as described in \citep{Essig:2018pzq}, 

\begin{equation}
t_c = \frac{150}{\beta}\frac{1}{r_s \rho_s \sigma_{\textrm{eff}}/m}\frac{1}{\sqrt{4\pi G \rho_s}}, 
\label{eq:collapse_time}
\end{equation}
where $r_s$ is  the scale radius in the NFW profile of the halo, i.e. the radius at which the logarithmic slope of the density profile reaches $-2$, $\rho_s$ is a scaled density related to the density at scale radius  and $\beta$ is a numerical factor that compensates for differences between the approximate fluid simulations and full N-body simulations of elastic SIDM in the long-mean-free-path regime \citep{Rocha12083025}.

\begin{figure}
    \centering
    \includegraphics[width=\columnwidth]{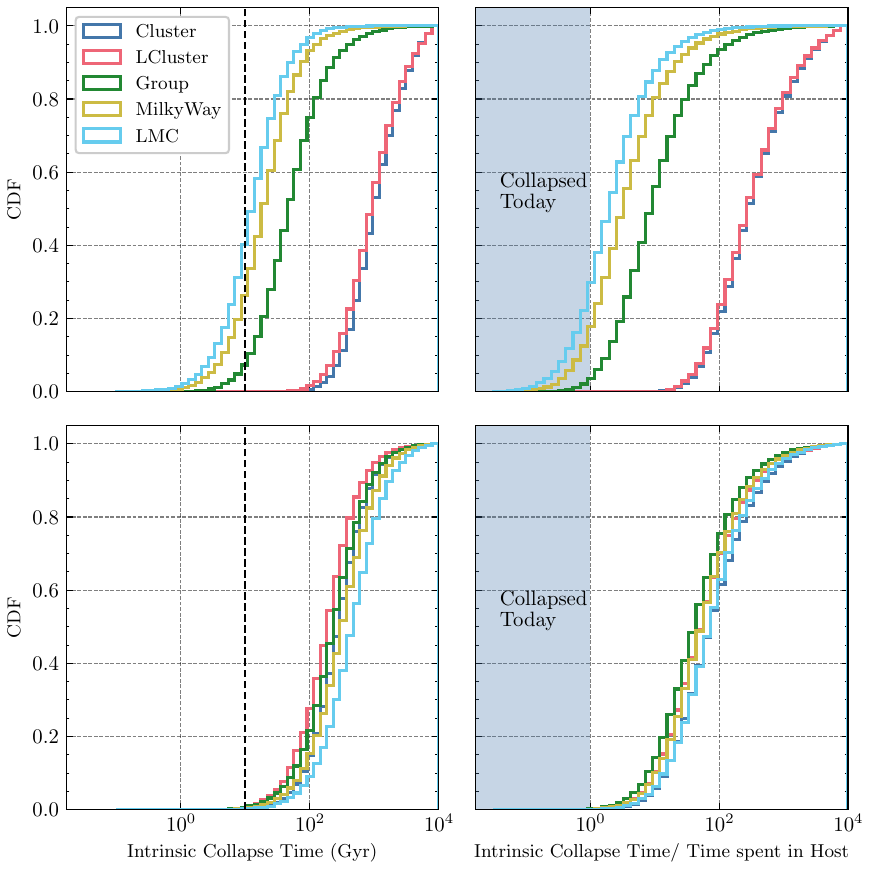}
    \caption{The cumulative distribution function for the intrinsic collapse time of subhalos that have survived to $z=0$ within their hosts for different host mass halos for the two cross-section models. The top panel shows the total collapse time and the collapse times scaled by the amount of time the subhalos have spent in their host for the YNY23 model, while the bottom panel shows the same results for the KTY16 model. The shaded region corresponds to halos that will have collapsed intrinsically since their infall on to the hosts.}
    \label{fig:collapse-time-cdf}
\end{figure}

\begin{figure*}
    \centering
    \includegraphics[width=\textwidth]{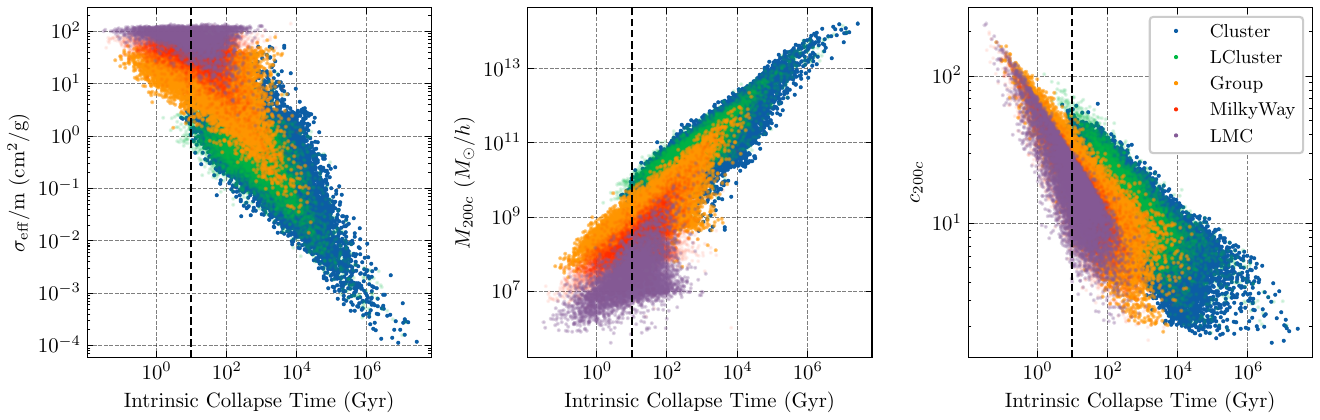}
    \caption{The dependence of intrinsic collapse times on infall properties for subhalos for the different host halos. The leftmost panel shows the $\sigma_{\rm eff}$ of the subhalos evaluated from the fiducial velocity dependent model as a function of the collapse times, the middle panel the mass of the subhalo at infall and the last panel the concentration. Subhalos for different hosts occupy different region of the property-collapse time space. Note that here we are plotting our entire subhalo population, without imposing the 300 particle cut at accretion. We do not include the poorly resolved subhalos in our main results.}
    \label{fig:collapse-time-dependence}
\end{figure*}

We use this scaling relation to estimate the collapse times for each subhalo in our simulation to get their timescales for collapse in the isolated scenario. We use the values of $r_s$ and $\rho_s$ for the subhalos at the time of their infall. These values have been obtained by fitting the NFW profile to the density profiles of the subhalos in the simulations at different snapshots and are provided with the Rockstar halo catalogs. These timescales then give us an estimate for the collapse time of halos in the scenario where they do not fall into the host. For the collapse timescales in tidal environment we semi-analytically solve the full gravothermal fluid equations at different points in the parameter space spanned by the subhalos in Section \ref{sec:fluid}. 

In Fig. \ref{fig:collapse-time-cdf} we plot the cumulative distribution of the intrinsic collapse times evaluated using Eq. \ref{eq:collapse_time} for both the interaction cross-sections given in YNY23 and KTY16 for each simulation suite. The right panel shows the  distribution of the collapse time scaled by the time spent by the subhalo in the host. In YNY23, which allows for high interaction cross-sections at low velocity scales, we find that for the low mass hosts, i.e. for the Milky Way, LMC  mass halos, more than $\sim 20$ percentile of all resolved subhalos have collapse times that are shorter than the time they have spent within their hosts. This implies that a large number of the subhalos within these objects are likely to have already gotten into a phase of rapid gravothermal core--collapse with significantly concentrated inner densities even in the absence of tidal stripping. For the Group mass host halos, this fraction is around $\sim 5$ percentile of subhalos. For the Cluster and L-Cluster suites we find that the fraction is nearly $\sim 0$ due to the low $\sigma_{\rm eff}/m$ at high velocities in the YNY23 model. These results imply that even without the effect of tidal truncation we should expect a significant fraction of collapsed substructure within low mass host halos simply because of the distribution of their inner densities and the velocity--dependent nature of the cross--section, that has higher interaction probability at lower velocities. A small fraction of the collapse of these objects are expected to be stalled due to tidal heating of the halos \citep{Zeng:2021ldo}. The dependence of the collapse times on different subhalo properties at accretion are demonstrated in Fig. \ref{fig:collapse-time-dependence} for the YNY23 model. For the model parameters in KTY16 we expect a negligible fraction of the subhalos to have collapsed intrinsically in the age of the universe across the simulation suites.

In the YNY23 model the differences between the high and low mass hosts have several contributing factors. Firstly, the mass function for subhalos is different for high and low mass hosts due to the resolution choices, in particular the mass function of the clusters and groups fall dramatically at values smaller than a few $10^{10} M_{\odot} h^{-1}$ and $10^9 M_\odot h^{-1}$ respectively. In principle these objects can also have a lot more low mass substructure, and the current distributions should be interpreted as a lower limit. It is essential to note that our percentiles show the fraction of collapsed subhalos that are resolved in the simulations. The subhalos we do resolve in the group simulations are quite massive and we find nearly $10$ percentile of them will collapse within Hubble time ($10$ Gyr). However, in addition to resolution effects, clusters and groups are also significantly younger than the rest of the hosts, as is evident in Fig. \ref{fig:subhalo-properties}, the average time spent by subhalos in a higher mass hosts is much smaller than the average time spent in low mass hosts, therefore the histograms for collapse times normalized by infall time into the massive hosts show even fewer collapsed percentiles.  These group subhalos with masses peaked at nearly $10^{9} M_\odot/h$, host dwarf galaxies that are well above the star--formation threshold and should allow us to study the stellar response of galaxies to core--collapse. Moreover the mass function of subhalos in Milky Way or smaller hosts is also suppressed at  masses higher than a few times $10^8 M_\odot/h$, due to dynamical friction, therefore these range of dwarfs are preferentially found in the tidal environments of group mass hosts.

\subsection{Collapse in Tidal environments}

So far we have estimated the intrinsic collapse time for the subhalo, i.e. the collapse time for these objects if they remained isolated through their lifetime. However, as a subhalo falls into a host several different factors can affect its core collapse time. \citet{Nishikawa190100499} showed that the core collapse timescales get accelerated due to tidal stripping of the outskirts of the subhalo. Tidal stripping essentially steepens the subhalo density profile outside its tidal radius due to the stronger tidal forces of the host, and this steepening accelerates the process of core collapse as it favours the outward heat transfer needed for it. 

Since we have a large sample of subhalos, we approach this problem in a statistical sense. We sample the parameter space of subhalo properties, primarily in $\rho_s$ or $r_s$ and the subhalo mass at infall, assume their profiles to be initially NFW and evolve them using the gravothermal fluid collapse model described below. We also tidally truncate their tails at first pericenter passage to simulate the effects of tidal stripping. Then we interpolate in this space to get the timescales for all subhalos in our simulation.
In particular we are following the approach taken by \citep{Nishikawa190100499}. Note that \citet{Zeng:2021ldo} showed from controlled simulations of single hosts that the effect of orbiting a potential adds a few more subtle effects, collapse can be decelerated depending on various effects such as tidal heating, and subhalo evaporation. Tidal heating, which is primarily a gravitational effect, heats the outer parts of the subhalo, which decelerates the process of core collapse to some extent, as the temperature gradient flattens. Subhalo evaporation also disrupts and delays the process of core collapse, however the effect is weaker if the cross--section is velocity dependent as the relative velocity in an interaction can be high. In a velocity dependent cross-sections the interaction cross-section between subhalo and host particles is suppressed by $v^{-4}$, therefore we don't expect significant effects of subhalo evaporation due to self-interactions. However, tidal heating can affect our collapse times too, we do not take it into account for this work.

\subsubsection{Gravothermal Fluid Model}
\label{sec:fluid}

To simulate an isolated SIDM halo, we use a semi-analytic model that was originally developed to study star clusters \citep{1980MNRAS.191..483L,1996NewA....1..255Q}, but has later been used to study SIDM halos in detail as well \citep{Gnedin:2000ea,PhysRevLett.88.101301,Balberg:2002ue,Pollack:2014rja,2011MNRAS.415.1125K}. It is computationally inexpensive to use compared to live, full N-body, simulations, and can resolve the inner regions of the halo as well as a wide parameter space. We treat the SIDM halo as a spherically symmetric gravothermal fluid, and the equations that describe such a fluid are;

\begin{equation}
    \frac{\partial M}{\partial r} = 4 \pi \rho r^2 
\end{equation}

\begin{equation}
    \frac{\partial (\rho v^2)}{\partial r} = \frac{-GM\rho}{r^2} 
\end{equation}

\begin{equation}
    \frac{\rho v^2}{\gamma -1}\left(\frac{\partial}{\partial t}\right)_M \rm ln \frac{v^2}{\rho^{\gamma-1}} = -\frac{1}{4\pi r^2} \frac{\partial L}{\partial r}
\end{equation}

\begin{equation}
    \frac{L}{4\pi r^2} = -\kappa \frac{\partial T}{\partial r} 
\end{equation}

where, $M(<r,t)$ is the mass enclosed within radius $r$ at time $t$, $\rho(r,t)$ is the local density, $v(r,t)$ is the one 1D velocity dispersion in the halo, $L(r,t)$ is the luminosity, $\kappa$ is the conductivity and $\left(\frac{\partial}{\partial t}\right)_M$ is the Lagrangian time derivative. Assuming the DM particle to be monoatomic, we set $\gamma = \frac{5}{3}$. The conductivity in the short-mean-free path regime is given by $\kappa_{\rm smfp} = \frac{3}{2}\frac{b \rho \lambda^2}{a m t_r}$, where $b\approx 1.38$, $a = \sqrt{\frac{16}{\pi}}$, $\lambda = \frac{1}{\rho \sigma}$ is the mean free path, $m$ is the DM particle mass and $t_r = \frac{\lambda}{a v}$ is the relaxation time. In the long-mean-free-path regime, the fluid equations require a parameter $\beta$ to calibrate them to SIDM N-body simulations. In this work, following \citep{Essig:2018pzq}, we set $\beta = 0.6$. The conductivity in the long-mean-free path regime can then be written as $\kappa_{\rm lmfp} = \frac{3}{2}\frac{\beta \rho H^2}{mt_r}$ where $H = \sqrt{\frac{v^2}{4 \pi G \rho}}$ is the gravitational scale height. We can then define conductivity in our equations as $1/\kappa = (1/\kappa_{\rm smfp} + 1/\kappa_{\rm lmfp})$. To solve the above equations numerically, we follow the procedure as used in \citep{Essig:2018pzq}. We solve the equations in terms of dimensionless quantities where the density and radius are scaled the NFW $\rho_s,r_s$, and the time is scaled by the dynamical time at the scale radius. The scaled quantities are given by,  $\rho = \tilde{\rho}\rho_s$, $M_0 = 4\pi r_s^3\rho_s$, $M = \tilde{M}M_0$, $\sigma/m = \tilde{\sigma}\left(\frac{4\pi r_s^2}{M_0}\right)$, $t = \tilde{t}\frac{1}{\sqrt{4\pi \textrm{G}\rho_s}}$ and other reference quantities are similarly written in terms of $\rho_s,r_s$. 

We discretize the equations and divide our halo into $150$ uniformly log-spaced concentric shells located between $\tilde{r}=0.01$ and $\tilde{r} = 100$. We then allow  a conduction time-step which allows heat exchange between the different shells. This pushes the halo out of hydrostatic equilibrium. Hence, to bring it back in equilibrium, we allow the radii of the shells to change, while keeping the density constant. The Knudsen number, $Kn$ is defined as the ratio of the mean free path of scattering, $\lambda$ and the gravitational scale height, $H$. When the mean free path becomes very small compared to the gravitational scale height the halo enters into core--collapse. We evolve our simulations using this process until the Knudsen number ($Kn$) of the innermost shell falls below $0.1$, which we define as the moment when the halo has core--collapsed.

\subsubsection{Modelling orbital evolution}

As noted before, we use a velocity-dependent cross section, to assign different effective cross-sections to different subhalos, depending on their maximum internal circular velocity. The Symphony suite gives us the complete orbits and history of the subhalos that survive till $z=0$. We find the infall time and the pericenter crossing time, where we define the pericenter time as the time taken to reach its first closest passage in the host after infall.

We treat the first pericenter passage as our only tidal truncation event, as this is when the tidal forces of the host are strongest and CDM simulations have found that this is the epoch when the subhalo loses most of its outer mass \citep{Nadler200108754}. We ignore evaporation and tidal heating effects which can also play a significant role in delaying core collapse if the host scale cross-section is large or if the intrinsic collapse time of the subhalo is similar to its pericenter crossing time \citep{Zeng:2021ldo}. To model tidal truncation in our fluid model, we change the density profile as follows (following \citep{Nishikawa190100499});

\begin{equation}
\rho(r>r_t) = \rho(r>r_t)\left(\frac{r_t}{r}\right)^{p_t}    
\end{equation}

where $r_t$ is the truncation radius and $p_t$ is a steep power-law cut-off in the density profile to model truncation. We set $p_t = 5$ \citep{2010MNRAS.406.1290P} in this work and show results for both $r_t = r_s,3r_s$. In a realistic scenario, the subhalos experience tidal effects of varying strength throughout their orbital history. In a previous work \citep{2021MNRAS.503..920C}, did include this effect with a tidal radius that depended on the distance from the centre of the host to constrain the self-interaction cross sections of various Milky Way satellite galaxies. However, we do not follow this approach since our simulation suites consist of thousands of subhalos with different orbital evolution histories, and we only wish to calculate the acceleration of collapse times due to tidal effects in a statistical sense.

Using the infall subhalo properties, we can calculate the dimensionless $\sigma_{\rm eff}/m$ and dimensionless pericenter crossing time, i.e. $\tilde{\sigma}_{\rm eff}/m$, $\tilde{t}$ for all the subhalos across all the simulations. There are thousands of subhalos across all simulations and it is computationally infeasible to evolve them individually using our fluid model. Therefore, we create a 2D grid in the $\tilde{\sigma}_{\rm eff}/m$, $\tilde{t}$ space and evolve subhalos for each grid point using the fluid equations. We can then assign collapse times for all subhalos through interpolation in this 2D space.

\subsubsection{Results}

\begin{figure*}
     \includegraphics[width=0.9\textwidth]{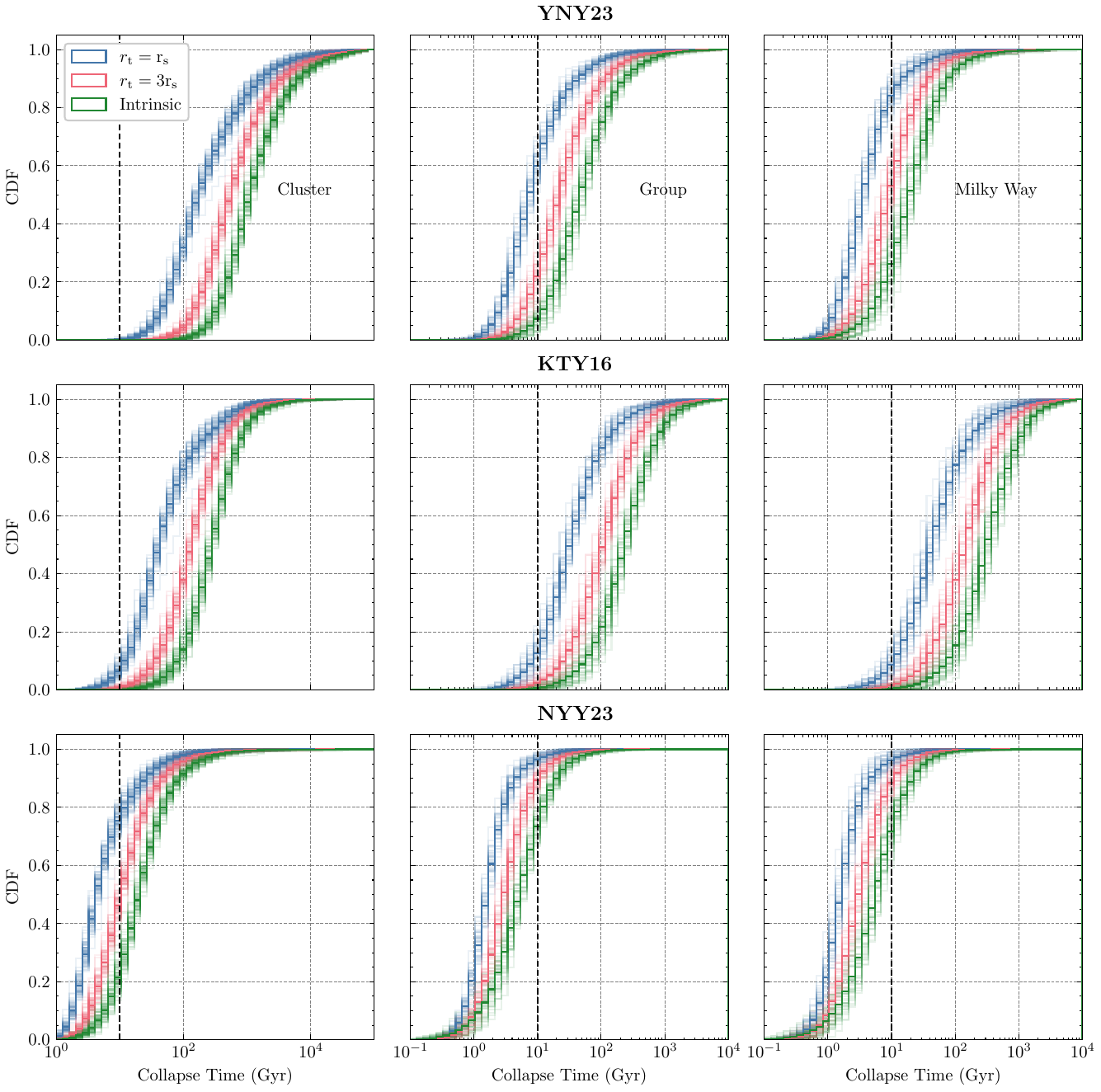}
    \caption{The cumulative distribution of collapse times evaluated using the fluid model in a tidal environment and their comparison to the intrinsic collapse scenario. For each color, the different curves correspond to separate host halos showing the scatter in the collapse times as a function of different evolution histories. The top panel shows the distribution found using the YNY23 model, and the bottom panel shows the distribution found using the KTY16 model. The blue and red colors correspond to a radius of $1 r_s$ and $3 r_s$ at which the tidal truncation is performed in the fluid simulations, and the green color corresponds to the case with no tidal truncation.}
    
    \label{fig:collapse-time-all-suites-scatter}

\end{figure*}

\begin{figure}
\centering
     \includegraphics[width=0.85\columnwidth]{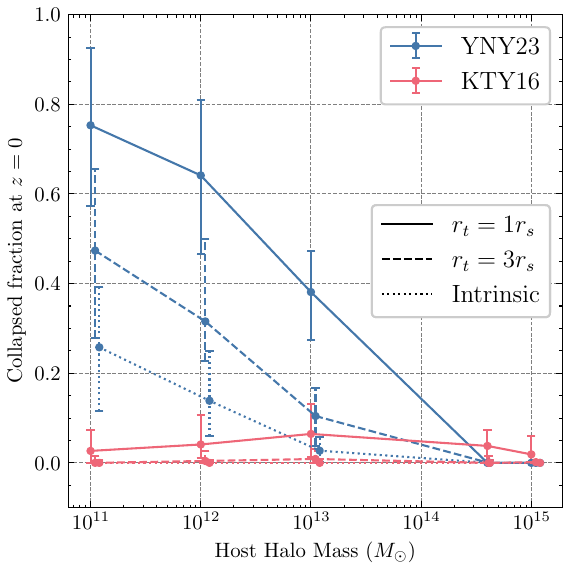}
    \caption{The fraction of subhalos that collapse by $z=0$ across all the host simulation suites. A subhalo is considered collapsed by $z=0$ if the collapse time is smaller than the time it has spent in the host halo after accretion. The collapsed fraction is shown for 2 different cross section models, YNY23 and KTY16 and for 3 different cases, tidal truncation at $r_s,3r_s$ and for no tidal truncation. For each host suite, there are multiple simulations in that suite. We have calculated the collapsed fraction individually for all the simulations, and the data point represents the median of those values, with the error bars showing the maximum and minimum collapsed fraction across all the hosts in that suite. }
    
    \label{fig:collapsed-fraction}

\end{figure}

We solve the dimensionless fluid equations for a 2D grid of 1200 uniformly log-spaced points in the $\tilde{\sigma}$, $\tilde{t}$ space. The former decides the strength of the self-interactions and the latter decides the time at which the profiles are truncated, to resemble the tidal stripping of the subhalo due to the host. We also solve the equations for a 1D grid of 40 uniformly log-spaced points in the $\tilde{\sigma}$ space without any truncation for scenarios where the subhalo has not reached pericenter by $z=0$. Note that for such subhalos, their collapse times in the host would be similar to their intrinsic collapse times as evaluated using Eq. \ref{eq:collapse_time}, which is expected as tidal effects are weaker away from the pericenter \cite{Zeng:2021ldo}. We solve these equations twice at each grid point for two different tidal truncation scales, with $r_t = r_s,3r_s$; the former representing a strong truncation event. The latter, larger truncation radius is closer to what is physically expected for low mass halos that have high concentrations and therefore very small scale radii, $r_s$ \citep{vandenBosch:2017ynq}. 

Given collapse times evaluated for the each point on the 2D grid, we can interpolate to get the collapse times for each individual subhalo using the $\tilde{\sigma}_{\rm eff}/m$ calculated from the YNY23 model or the KTY16 model and the $\tilde{t}$ from the pericenter times. Fig. \ref{fig:collapse-time-all-suites-scatter} show the distribution of collapse times in the presence of tidal truncation for each subhalo in our grid. As expected the collapse is accelerated in the tidal environment of the halo in comparison to the intrinsic collapse time. The top panel shows the results for YNY23 and the bottom panel for KTY16. In YNY23, for tidal truncation of the profile at $r_s$, the halos with collapse time less than $10~\rm Gyr$ form nearly $90$ percentile of the total population in Milky Way and around $60$ percentile of Group like systems whereas, it is nearly negligible in cluster mass halos. Assuming a more realistic tidal truncation at $3r_s$, we note that this fraction of halos falls to about $50-60$ percentile in Milky Way and smaller hosts and closer to $30$ percentile in the group mass hosts. We also note that there exists a significant halo to halo scatter in the distribution of collapse times since we have multiple host halos in each simulation suite. In comparison with previous work, \cite{2023ApJ...949...67Y} perform full N-body simulations on a single Milky Way like system for the cross-section that we call the YNY23 model as mentioned before. Applying the the same mass cuts on the subhalos masses as theirs we find that the percentile of collapsed objects within the Milky Way in our work agrees with their results within the halo--to--halo scatter, as can be seen in Fig. \ref{fig:collapsed-fraction}.  The middle row of the same figure shows the cumulative distribution of collapse times using the KTY16 model. Tidal truncation in the host environment enhances the collapsed fraction of subhalos even in the KTY16 models. For the realistic case of a truncation at $3 r_s$ nearly $5$ percentile of Milky Way and Group subhalos are expected to have collapse times within the age of the universe. Therefore even the current stringent constraints that allow effective cross-sections of about 2 $\cmg$ at low velocities, we can expect $5-10\%$ of subhalos to collapse in tidal environments. We again emphasize that the fraction of subhalos in our results is to be interpreted as the fraction of the subhalo population above our resolution limit, and not as the entire subhalo population, which includes lower mass subhalos as well that we do not resolve or take into account. 

In the bottom panel of the same figure we also show the same results for a velocity dependent cross section with parameters explored as in \cite{2023arXiv230601830N} we refer to this as NYY23. In this model, the $\sigma_0$ is the same as YNY23, however the $v^{-4}$ cut-off scale is much larger at $w = 120 \rm km/\rm s$; This allows for high cross--sections at high velocities along with them being high at dwarf scales, and naturally leads to more gravothermally collapsed subhalos in groups and cluster scales as compared to the previous two models. These parameters have been explored because they are consistent with the dense substructure in the strong lens galaxy SDSSJ0946+1006 \citep{Minor:2020hic} and simultaneously with low concentration dwarfs. The current functional form of the velocity--dependent cross-section is such that to predict high cross-sections at both these mass scale it allows for cross-sections of $\sigma_{\rm eff}/m>10 {\rm cm^2/g}$ on Milky Way scales as well, which are hard to explain. But such behaviour is interesting to explore in a model free way and to see the predictions of the velocity dependence on the subhalo collapse function. Fig. \ref{fig:collapsed-fraction} shows the cumulative fraction of collapsed subhalos that collapse within their time in the host and the scatter in the fraction.

\section{Discussion}
\label{sec:discussion}
\subsection{Comparison to Isolated halos}
\label{sec:isolated_halos}

\begin{figure*}
\centering
    \includegraphics[width=0.75\textwidth]{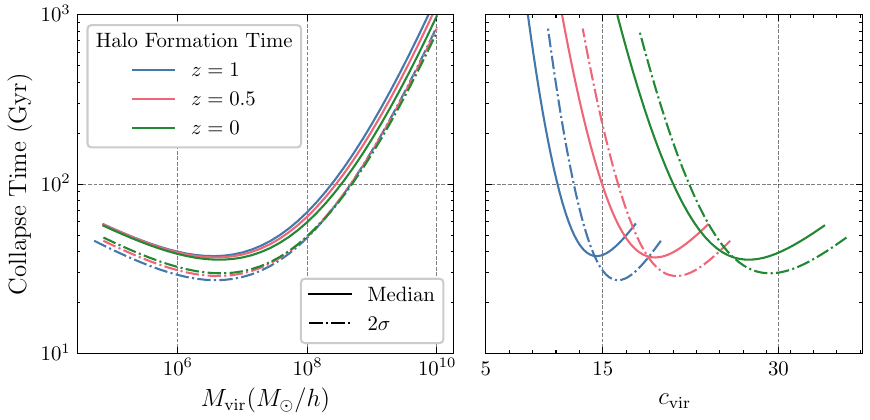}
    \caption{The two panels show the collapse times for isolated halos in the universe as a function of mass and concentration as expected from the mean concentration--mass relation in $\Lambda$CDM. The halo formation times are taken to be z=0,0.5,1. Isolated halos following the mean relation and within $2\sigma$ of the mean are not expected to collapse within Hubble time.}
    \label{fig:isolated_collapse_time}
\end{figure*}

\begin{figure*}
\centering
    \includegraphics[width=0.75\textwidth]{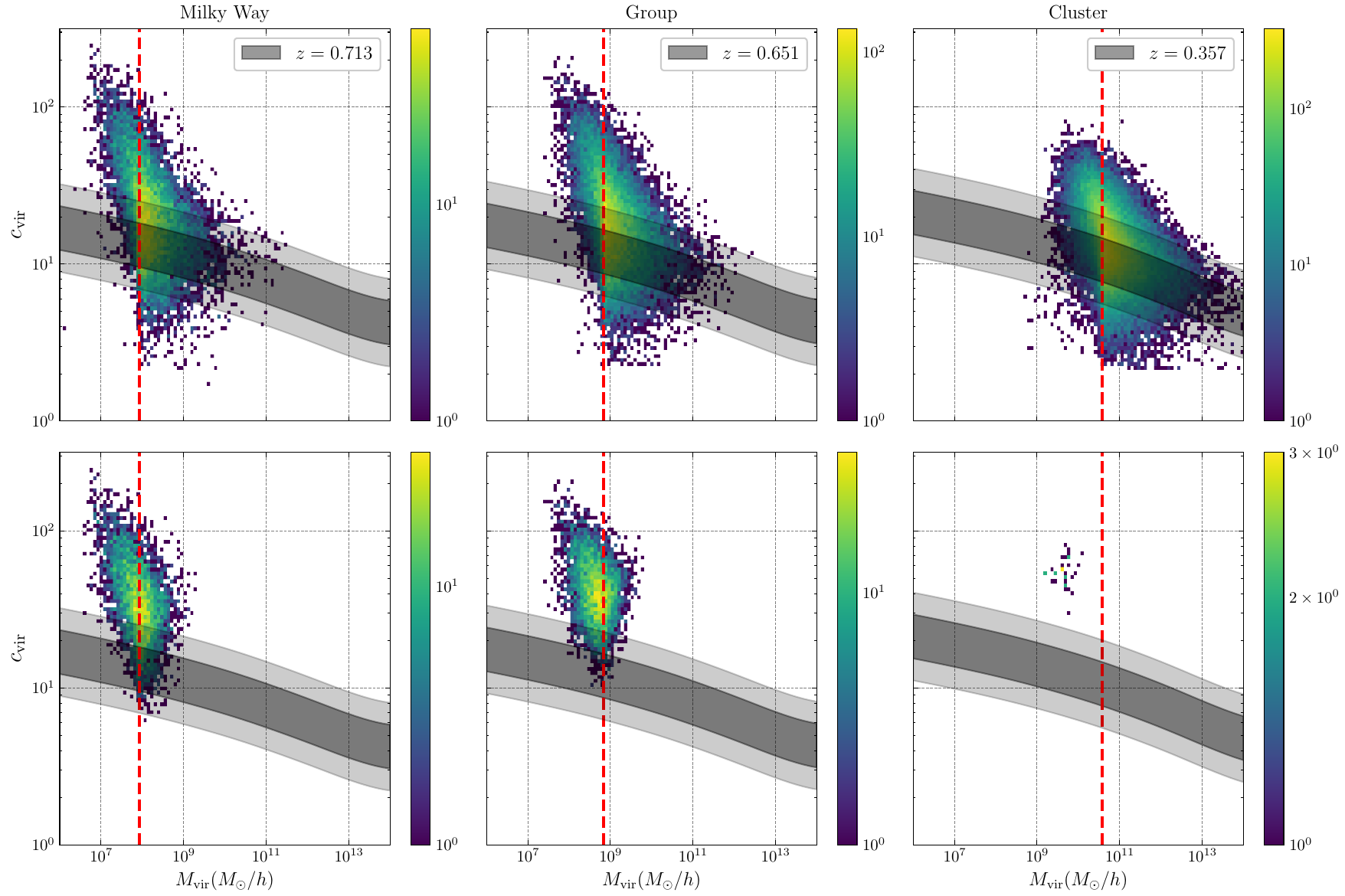}
    \caption{The top panel shows the concentration--mass relation for the subhalos of Milky Way, Group and Cluster suite. The gray bands show the concentration mass relation \citep{Ishiyama190703642} at the median redshift of infall for subhalos in that particular suite. The bottom panel shows the concentration--mass relation for the subhalos of Milky Way, Group and Cluster suite that have intrinsic collapse times lesser than $10$ Gyr as evaluated using the YNY23 model. Note that here we are plotting our entire subhalo population, without imposing the 300 particle cut at accretion. We do not include the poorly resolved subhalos in our main results. The red line roughly lies at the 300 particle mass for each suite.} 
    \label{fig:subhalo-properties-collapsed}
\end{figure*}

Here we explore the collapse times for isolated objects that do not live as subhalos of a more massive hosts. As we want to explore core--collapse, we only show results for the case of YNY23. Figure \ref{fig:isolated_collapse_time} shows the distribution of collapse times for isolated halos as a function of their concentrations, masses assuming a standard concentration--mass relation at a given redshift from \cite{2021MNRAS.506.4210I} and a scatter of $0.14$ dex \citep{Wechsler2002}. 

Assuming the halos to have an NFW profile, we calculated its maximum circular velocity which is used to infer its effective cross section. We find that for halos that follow the mean $c-M$ relation or have concentrations higher than the mean, but within $1\sigma$ or even $2\sigma$ scatter of the mean, the core collapse time is greater than the Hubble time. This agrees with previous results that the intrinsic collapse time for most cases is much longer than the age of the universe, hence isolated halos are expected to be in their core forming phase. This is puzzling when we consider the fact that a large fraction of objects that end up as subhalos do in fact end up collapsing within a Hubble time even without tidal effects. This implies that subhalos (at least in simulations) are somehow biased towards higher concentrations and are more likely to collapse. Such an effect though is expected in hierarchical CDM structure formation. A correlation between halo concentration and density is known as assembly bias \citep{Wechsler2002, Wechsler:2005gb, Gao:2006qz, Dalal:2008zd}, at low masses in particular higher concentration halos are more clustered than low concentration ones. Part of this effect is simply due to tidal effects on halo profiles in dense environments that tend to steepen their mass distribution \citep{Mansfield:2019ter}. In figure \ref{fig:subhalo-properties-collapsed} we show the distribution of the subhalo populations in the $c-M$ plane at the time of their infall, for three of our simulation suites. We also show the $1\sigma$ and $2\sigma$ bands for the concentration mass relation for all halos in the box at the mean accretion redshift for the subhalos in that suite. We find that the subhalos can often have a much larger scatter than the mean relation. The distribution of subhalos with a intrinsic collapse time less than $10\rm Gyr$ is shown in the bottom panel. We defer a detailed exploration of the particle profiles in different environments and the effect of self--interactions to a future study.
\subsection{Observational Implications}
\label{sec:obs_imp}

Our analysis shows that a significant fraction of subhalos are expected to collapse within a Hubble time, in this section we assess the observational implications of this population. Several studies have used the observed central densities of Milky Way satellite systems to constrain models of dark matter. Tracer stars within dwarf galaxies are typically used to estimate a dynamical mass within some radius near the central region of the object, and a density is inferred. 
In this work we track the evolution of the central density as a function of time in our fluid simulations, we can therefore estimate the distribution of densities of the existing subhalos in all the suites. 
\begin{figure*}
    \includegraphics[trim={0 0.5cm 0 0}, width=1.2\columnwidth]{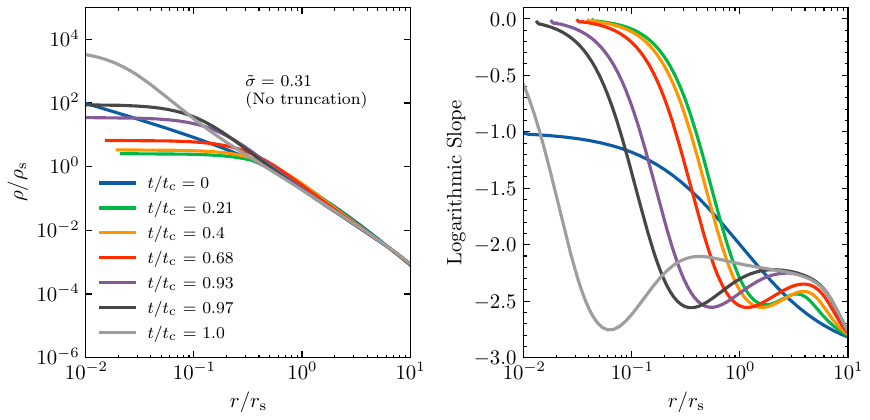}
    \includegraphics[width=1.2\columnwidth]{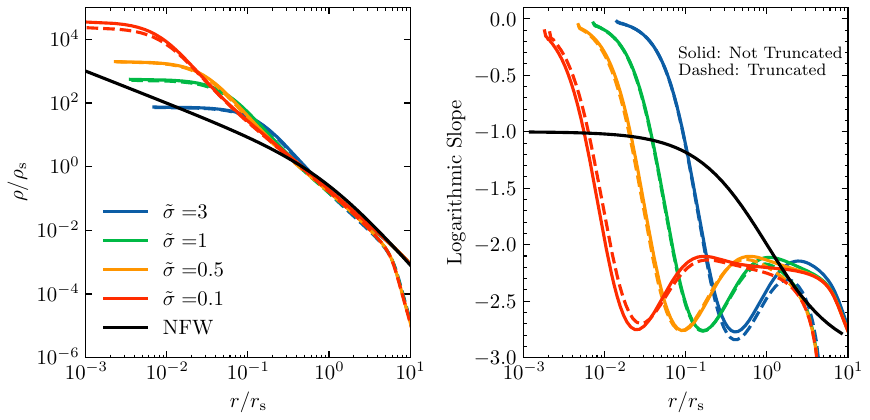}
    \caption{The evolution of the density of the halo and its logarithmic slope from the fluid simulations during core--collapse. The top panel shows an example of the density and slope evolution for a Milky Way subhalo, as a function of time after the core--expansion phase through core--collapse. The time steps are represented as a fraction of the collapse time ($t_{\rm c}$), which is chosen as the time when the Knudsen number of the innermost shell falls below 0.1, in principle the subhalos can already be in the collapsing phase before $t_c$. The bottom panels shows the final profiles (at $t_c$) for various different subhalos in the sims, depending on the subhalo parameters they can be at different stages of core--collapse at the final time $t_c$. Note that at $t_c$ the slope around the scale radius approaches a constant value of $\sim -2.1$ and, in general, with the progression of core--collapse the size of that region increases as can be seen in the top panel. }
    \label{fig:density_slope}
\end{figure*}

Firstly we study theoretically the evolution of the halo in the fluid simulations. The top panel of Fig. \ref{fig:density_slope} shows the solutions to the fluid equations for a typical Milky Way subhalo which has a median scaled $\tilde{\sigma} = 0.31$ in the YNY23 model. The left panel shows the evolution of the density profile and right panel shows the evolution of the logarithmic slope, $\Gamma=\frac{d{\rm log}\rho}{d{\rm log} r}$. The plots shows the evolution post the core--expansion phase when the halo has begun to core collapse, therefore one can see the core shrinking with time. There are several features we note here, firstly, the density values in the innermost bins within the scale radius starts diverging significantly from NFW, becoming larger than the expected value at late times. We note that a core--like feature continues to exist in the smallest radius bins, where self--interactions are the strongest. In the top right panel of the same figure, showing the logarithmic slope of the profile, we note that right outside the inner most bins where the core is,  the log--slope  quickly falls off to a maximum negative value, this is where the particles from the core essentially accumulate, the slope becomes less steep beyond that and approaches a value close to isothermal $\approx -2.1$. The profiles are eventually approaching a cored isothermal sphere with time and such a behavior of the slope profile is expected (see Fig. 4.7 in \cite{binney_tremaine_2008} 4th edition). The bottom panel of Fig. \ref{fig:density_slope} shows the final profiles for different values of $\tilde{\sigma}$ when the simulation is ended). This corresponds to the profiles at the point when we terminate the fluid evolution as $Kn$ falls below $0.1$. The dashed line in the bottom panel show the profiles when the outer region is tidally truncated, with a truncation radius of $3 r_s$. The overall features of the evolution remain the same on tidal truncation, except that the evolution is accelerated. 

\begin{figure*}
    \centering
    \includegraphics[width=0.75\textwidth]{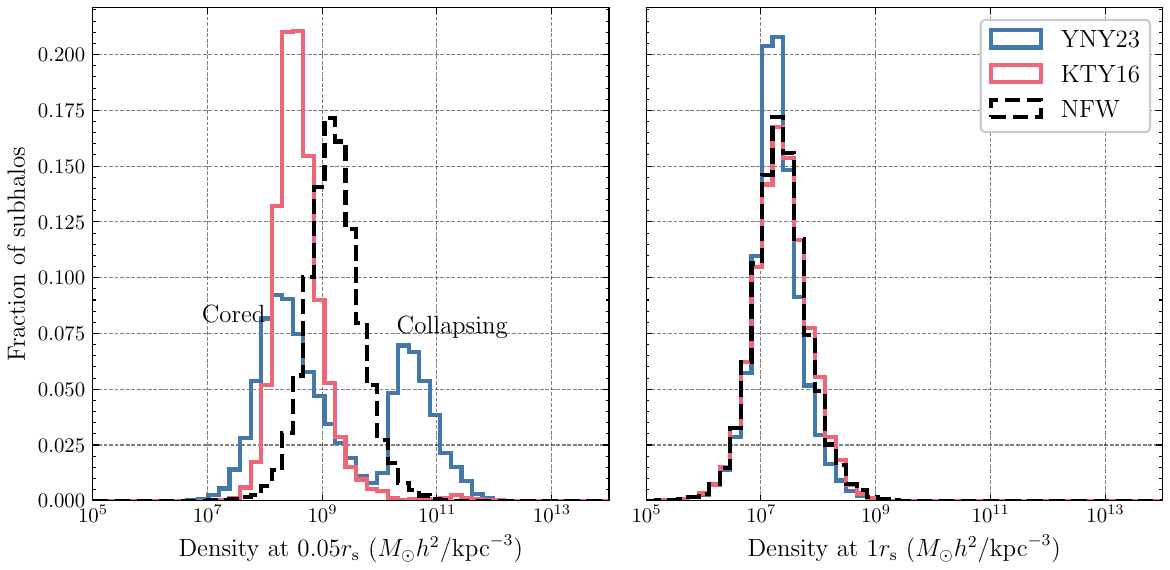}
    \caption{The left and right panels show the distribution of densities at 0.05 and 1 $r_s$ respectively of the Milky Way subhalos when they are evolved with the fluid simulations. It should be noted that for the collapsed subhalos, the value of density is the one calculated at the onset of core collapse, which is when we terminate the evolution, and not at $z=0$; hence they should be treated as lower limits for their densities today. The black dashed curves shows distribution of densities at the same radius for NFW profiles with parameters at the time of infall of the subhalo.}
    
    \label{fig:subhalo-density}

\end{figure*}

Measuring the slope and densities of subhalos at several points is naturally a very strong probe of core--collapse. While we see that the absolute value of the density at the scale--radius does not change dramatically from CDM, the slope becomes significantly different. The strong steepening to slopes steeper than $-2$, is not expected in the presence of baryons only even if we account for adiabatic contraction. In Fig. \ref{fig:subhalo-density} we show the histogram of the density in our Milky Way simulations deep inside the subhalo and at its scale--radius. These histograms correspond to the more conservative case where we assume tidal truncation occurs at $3 r_s$. The blue and red curves correspond to the YNY23 and the KTY16 models respectively, the black dashed curve corresponds to densities expected from NFW, considering the distribution of their concentrations at infall. The distribution of densities only begins differ appreciably from the CDM case deep inside the scale radius. In the runaway core--collapse phase the density near the center rises  nearly exponentially with time, this leads to a bi-modality in the central densities of subhalos in the presence of self--interaction, i.e. the densities for the population of collapsed subhalos separate out quickly from the uncollapsed population due to the small time spent in the intermediate valley region as can be seen in the left panel of Fig. \ref{fig:subhalo-density} for YNY23 where a large number of low mass subhalos collapse. Note that even though most halos in the KTY16 model are expected to have cores in the inner bin, there is a tail that mildly extends to the collapsing region. In either case the distribution of inner densities of subhalos are distinct from NFW.
To reiterate, we differentiate between "collapsed" and "non-collapsed" objects based on the Knudsen number, the densities of the collapsed objects are lower limits as we stop the simulation when the Knudsen number reaches the value chosen for collapse, in reality they may be higher at the time we observe them.

\begin{figure}
\centering
     \includegraphics[width=0.8\columnwidth]{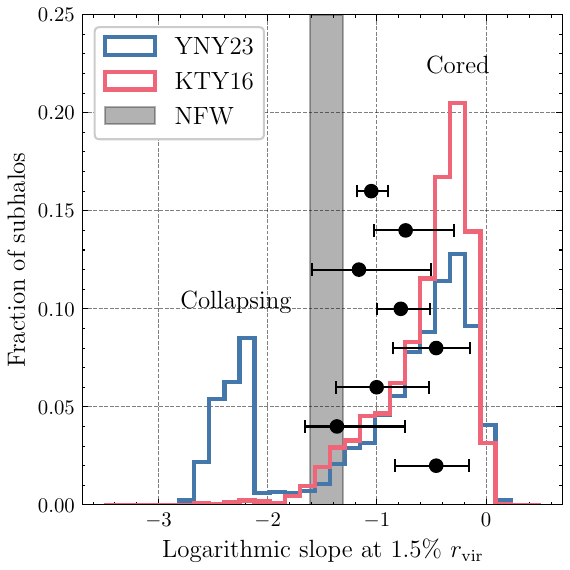}
    \caption{Comparison of log--slope distribution with data. The blue histogram shows the distribution of logarithmic slopes at $1.5 \% r_{\rm vir}$ of the Milky Way subhalos when they are evolved with the fluid simulations using the YNY23 model. The red histogram shows the same distribution using the KTY16 model. These subhalos are also expected to have $M_\ast/M_{\rm halo}>10^{-5}$. The black data points are the estimated logarithmic slopes at  $1.5 \% r_{\rm vir}$ for the classic and bright dwarfs in the Milky Way by \citet{Hayashi:2020jze}. The gray band represents the expected NFW slope values at 1.5 $\% r_{\rm vir}$ as derived from dark matter only simulations \citep{2016MNRAS.456.3542T}. The black data points correspond to the following dwarf satellites from bottom to top; \textit{Fornax, LeoI, LeoII, Sculptor, Carina, Ursa Minor, Sextans and Draco.}}
    \label{fig:log-slope-observation}.

\end{figure}

In Fig. \ref{fig:log-slope-observation} we compare the slope distribution to data for estimated slopes of 8 dwarf spheroidal (dSph) galaxies. We use the data quoted in \cite{Hayashi:2020jze}. As in \cite{Hayashi:2020jze}, we compare the observations to the the slopes at $1.5\%$ $r_{\rm vir}$ measured from our Milky Way suite subhalos. To match the sample of subhalos to data, we only use bright satellites with $M_{\ast}/M_{\rm halo}>10^{-5}$, where we have used the \citet{Moster:2012fv} stellar to halo mass relation to get a stellar mass for our subhalos using their virial mass at infall. The blue and red histogram show the distribution of slopes in our simulation for YNY23 and KTY16 respectively, the grey band shows the slopes expected for NFW halos \citep{2016MNRAS.456.3542T} and the black points are the data. We note that we don't see any classical dwarfs that appear to be in the phase of core--collapse, most of them in fact show a preference for a core.

One issue that we note in our work is that we primarily use the $\rho_s$ and $r_s$ measured in the Rockstar subhalo catalog to estimate NFW profiles and run the fluid simulation on those. In principle we could also take the particle profiles from the simulations themselves and they may have significant differences from NFW on a halo to halo basis. However, we expect that on average our results will hold, as NFW has been known to be a good fit, prior to infall and tidal effects for halo profiles. Further we use a 300 particle limit to define our halos at $z=0$, as well as at accretion, this limit is chosen based on the convergence tests performed in the original Symphony simulations suite. These simulations show that the subhalo mass functions and accretion histories are converged to better than $10\%$ using such resolution cuts, however we note that the profiles, in particular the concentration convergence is significantly more complicated to define and depends on the force-softening of the simulations \citep{Mansfield:2017}. We do not resolve the convergence issue in this paper, and use the spread that the simulations provide us, but in principle it will be useful to simulate a larger sample of higher resolution zoom--in simulations at the observable limit of low mass galaxies.  

In this work we do not take into account any baryonic physics, or the effects of SIDM starting from the initial perturbation peaks, both of which are expected to create differences from our current inferences, however we do not expect the trends to change dramatically. In the latter case it has been shown that fluid simulations that begin assuming NFW--like profiles reproduce similar trends as a full SIDM N-body treatment with a scaling factor in the conductivity to to match N-body simulations\citep{Koda11013097, Essig:2018pzq, 2023arXiv230608028Z}, we include this in our work. However, it is still worth developing simulations that will take into account the full evolution in SIDM \citep{2023ApJ...949...67Y} and running it on large cosmological boxes. 

With regard to baryonic physics, in general it has been seen in several works that including a baryonic potential can lead to a regeneration of a cusp in halos during the core--expansion phase \citep{Kaplinghat:2013xca, Sameie180109682, Sameie190407872, Robles190301469}, this can further lead to a more accelerated core--collapse. It has also been noted in \citep{Jiang:2022aqw,2023arXiv230608028Z} that semi-analytic SIDM models, that include the effects of baryons as well, lead to a more diverse halo response to the baryons than in CDM or SIDM without baryonic effects.

\section{Conclusions}

In this paper we have studied a large statistical sample of subhalos that live in a range of host halos extending from LMC mass to cluster masses to estimate the probability that they will core collapse during their lifetime if dark matter was self-interacting. We use the Symphony-Suite of simulation that are a compilation of data products from zoom--in simulations of 39 LMC, 45 Milky Way, 49 Group and 33 low and 96 high cluster mass host halos. We extract the evolutionary properties of all surviving subhalos at $z=0$ for each of these suites and use these to run fluid simulations of SIDM to predict the evolution of their internal density profiles in the tidal environment of the host. In SIDM all halos are expected to evolve from a core-expansion phase to a core--collapse phase, the time-scale for the transition collapse is expected to be low for high interaction cross-sections and in the presence of tidal truncation. In this work, we find the full distribution of collapse times for the subhalo population above our resolution limit in different hosts.

We evaluate, i) the intrinsic collapse times for each subhalo using properties pre-infall and ii) the timescales for collapse in a tidal environment, once these subhalos fall into their hosts, we tidally truncate their profiles at $r_s$ and $3 r_s$ at the time of their pericenter passage. We find that a significant fraction of the resolved subhalos that live in Milky Way, LMC or group like environments are expected to collapse within the Hubble time and within the time that they spent in their hosts in a model like YNY23. Even if we consider intrinsic collapse only, nearly $30$ percentile of the subhalos in Milky Way are expected to have collapse times less than $10$ Gyr, with a significant halo--to--halo scatter. Adding tidal truncation accelerates core--collapse in subhalos due to the earlier onset of outward heat transfer from the cored centre setting in faster. In Milky Way like systems a conservative tidal truncation can lead to collapse times less than $10$ Gyr for nearly $50$ percentile of substructure. Full cosmological SIDM zoom--in simulations of a Milky Way like host have been performed by \cite{2023ApJ...949...67Y}. Our percentile of collapsed objects (objects with collapse time smaller than time since accretion) for Milky Way like systems is consistent with the full N-body treatment within scatter.

We find that the fraction of collapsed substructure is larger in lower mass hosts like the LMC and Milky Way compared to groups and clusters. This is expected given the resolution limit of the simulations, and the fact that high concentration, low mass halos are typically collapsing. Despite that it is important to note that the collapsed fraction in groups is quite high for a model like YNY23, and also in KTY16 when tidal effects are taken into account. The distribution of satellite masses also shows that $10^9$ solar mass objects dominate the group environments and are rarer in Milky Way, LMC environments, therefore groups and clusters are good laboratories to study these objects in tidal environments. In principle there can be more collapse within group subhalos, if the cross-section of interaction is allowed to be higher at the halo mass scales where the mass function for group subhalo peaks, as explored in \citep{2023arXiv230601830N}. However, the current form of the velocity-dependent cross-section (\ref{eq:sigma_t}), does not allow this naturally, without simultaneously significantly increasing the cross-sections at Milky Way mass scales beyond the allowed parameter space constrained by data.

From an observational point of view the density at several points around the scale radius of the subhalos, or alternatively the slopes of the density distribution can be a robust probe of the nature of self--interactions. The observed central densities of Milky Way dwarf satellites that are measured at a few $100$s of parsec scales are typically probing the scale radius region. The distribution of densities however show more prominent differences as we move to smaller radii. Deep inside the densities begin to diverge away from NFW and create a bimodal distribution with one population having reached very high densities due to exponential increase in densities over a short timescale during core--collapse and another population in the coring phase. Comparing with slopes of classical dwarfs measured at $1.5\% r_{\rm vir}$, we find that the simulations with YNY23 model parameters predict too many collapsed objects  at these mass--scales, currently we don't see any evidence for a steepening of the slopes in the inner region from data. Assuming dark matter to be self interacting with a velocity dependent cross section model such as the one used in this work, the KTY16 model is more consistent with what is observed. 

Several studies have explored the impact of dark matter self-interactions on subhalos \citep{ DOnghia0206125, Vogelsberger2011, Dooley160308919, Rocha12083025, Zavala12116426, Robles190301469, Nadler200108754, Nadler:2021rpo, Bhattacharyya:2021vyd}, however most of these studies have been confined to cross-sections where core--collapse is not expected to occur. At higher cross-sections, where core--collapse can occur the first studies of full cosmological object was done by \citep{2023ApJ...949...67Y} and \citep{Nadler:2022dvo} for a single Milky Way and Group system. Our fiducial model parameters corresponds \citep{2023ApJ...949...67Y}. Our analysis is consistent with their results when the halo to halo scatter and the resolution limit is taken into account. However, in self-interacting dark matter simulations extending the cross-section to very high values can be significantly challenging, most simulation methods need to restrict having one interaction per time step, and assumptions about convergence still remain to be tested in the high interaction regime. While simulations methods are being developed and extended to include core--collapse, $\Lambda$CDM can give us important insights into the distribution of satellite properties. 

This work shows that given the distribution of properties of halos due to structure formation in CDM cosmologies, a range of satellites, including those that are bright and massive are expected to be in a phase of core--collapse particularly those that live as subhalos in hosts ranging from LMC to cluster mass objects. These collapsed systems should leave signatures in strong-lensing studies, weak--lensing profiles of satellites and also additionally on their galaxy properties like their sizes and luminosities. In future it will be useful to explore if such relations, such as the galaxy--halo size relation, luminosity relations also put constraints on the nature of self--interactions between dark matter particles due to the signatures left behind by the different stages of evolution.

\section*{Acknowledgements}
We thank Hiroya Nishikawa, Zhichao Carton Zeng, Daneng Yang, Ethan Nadler, Arka Banerjee, Phil Mansfield and Dhruv Hukkeri for useful discussions and help with the Symphony zoom--in simulations. We thank Yiming Zhong for providing help with the fluid simulation code. We  received support from the PARAM Brahma supercomputing facility at IISER Pune, which is part of the National Super Computing Mission under the Government of India.



\bibliographystyle{mnras}
\bibliography{references1, references2}





\bsp	
\label{lastpage}
\end{document}